\documentclass[prb,twocolumn,showpacs,superscriptaddress]{revtex4}
\usepackage{graphicx}
\usepackage{amsmath}
\usepackage{amssymb}
\usepackage{color,ifpdf}

\begin{document}

\title{Crossover from  diffusive to strongly localized regime in two-dimensional systems}

\author{A. M. Somoza}
\affiliation{Departamento de F\'{\i}sica,
     Universidad de Murcia, Murcia 30.071, Spain}
\author{J. Prior}
\affiliation{Departamento de F\'{\i}sica Aplicada,
     Universidad Politécnica de Cartagena, Cartagena 30.202, Spain}
\author{M. Ortu\~no}
\affiliation{Departamento de F\'{\i}sica,
     Universidad de Murcia, Murcia 30.071, Spain}
\author
 {I. V. Lerner}
 \affiliation{School of Physics and Astronomy, University of
Birmingham, Birmingham B15 2TT, UK}

\begin{abstract}
We have studied the conductance distribution function of two-dimensional disordered non-interacting systems in the crossover regime between the diffusive and the localized phases. The distribution is entirely determined by the mean conductance, $\langle g\rangle $, in agreement with the strong version of the single parameter scaling hypothesis. The distribution seems to change drastically at a critical value very close to one. For conductances larger than this critical value, the distribution is roughly gaussian, while for smaller values it resembles a log normal distribution. The two distributions match at the critical point with an often appreciable change in behavior. This matching implies a jump in the first derivative of the distribution which does not seem to disappear as system size increases. We have also studied $1/\langle g\rangle $ corrections to the skewness to quantify the deviation of the distribution from a gaussian function in the diffusive regime.
\end{abstract}
\pacs{71.23.-k, 72.20.-i}
\maketitle

\section{Introduction}

The distribution function $P(g)$ of the conductance of disordered systems
is quite well understood in the metallic diffusive regime. It has a Gaussian shape \cite{AKL86} and so its first and second moments are {sufficient} to describe it. The second cumulant (irreducible moment) is a {universal constant \cite{Al85} of order $e^2/h$} for a diffusive system of any dimensionality {which explains the term} \textit{universal conductance fluctuations}.

{On the other hand, the} distribution function $P(g)$ in the localized phase is  {known to be} log normal in one-dimensional (1D) systems \cite{Ab81}, where the metallic regime does not exist, and recently {has} been found to be a Tracey-Widom distribution for the strongly localized phase in two-dimensional (2D) systems \cite{SO07}.

The crossover between the diffusive and localized regimes is possible both in quasi-1D and in 2D systems.  Extensive studies \cite{GS01,GM02} of quasi-1D systems have shown the distribution function to be independent of the system details, with the average conductance being the only scaling parameter. This agrees with the  one-parameter scaling hypothesis  formulated for the mean conductance  \cite{AALR} and later extended to the entire conductance distribution function \cite{Sha87}. On the insulating side of the quasi-1D crossover, $P(g)$  has been found to be  essentially a ``one-sided" log-normal distribution \cite{GS01} in agreement with analytical studies of strictly 1D systems. \cite{Ab81}

The validity of the one-parameter scaling hypothesis for the conductance distribution function has also been
 studied in the localized phase of 2D systems \cite{SO07,PS05,SA04} where $P(g)$ is now understood almost as well as in the diffusive phase. However,
the crossover regime between the diffusive and the localized
phases, where the localization length is of the order of the system size, is poorly
understood in 2D systems. The numerical investigation of this regime is the focus of this paper.

The  conductance distribution can be described by its cumulants,
which have been found in the diffusive regime to have the form \cite{AKL86}
\begin{equation}\label{1}
\langle g^n\rangle_c\propto  \langle g\rangle ^{2-n},\quad n<g_0,\quad \langle g\rangle\gg 1
\end{equation}
Here $g$ is the dimensionless conductance (measured in units of $e^2/h$),  the brackets $\langle \dots \rangle  $ stand for the ensemble averaging over disorder, $\langle \dots \rangle _c$ denotes a cumulant and $g_0$ is the mean conductance at the scale of the elastic mean free path $\ell $.
It has been shown  \cite{AKL86} that there are no perturbative corrections to the second cumulant (variance). Therefore, in the absence of non-perturbative corrections it would remain universal (of  order $1$)  for any value of $\langle g \rangle  $. Since in the regime of  strong localization,  $\langle g \rangle \ll1$, the variance should eventually decrease, and the non-perturbative corrections must exist. 

In the absence of a sharp transition between localized and delocalized states in $2D$ systems, it is possible that such non-perturbative contribution to the variance can be numerically traced already at the threshold between the diffusive and localized regimes where $\langle g \rangle \sim1$. As for the higher cumulants, in the metallic and crossover regime one can assess numerically only the third cumulant of the distribution, which represents the leading deviation from the Gaussian. Equation Eq.~(\ref{1}) shows that it
is proportional to $1/\langle g\rangle$. The constant of proportionality has been calculated diagrammatically\cite{vRLAN}
and the result for quasi 2D systems is 
\begin{equation}\label{2}
\langle g^3\rangle_c= -0.0020 \langle g\rangle ^{-1}.
\end{equation}

On the other hand, one can study numerically how the distribution function changes through the crossover regime from the almost Gaussian for $ \langle g \rangle >1$ to almost log-normal for $\langle g \rangle <1$. Analytical studies predict also the appearance of the log-normal tails in the crossover regime \cite{AKL86} which signify the emergence of pre-localized states \cite{FE95}. However, their statistical weight might be rather small in order to trace them numerically.

Thus the aim of this paper is two-fold. First, we want to study systematically the full distribution function
$P(g)$ in the crossover regime and check if the scaling hypothesis extended to the full distribution
applies in this region. Second, we want to obtain the corrections to the variance and the leading
contribution to the third cumulant as a function of $1/\langle g\rangle$ as we move away from the
deep metallic regime into the crossover region. 

In the next section, we describe the model and the numerical procedure. In section III,
we obtain the conductance distribution function for a wide variaty of values of the disorder
and the system size and present the results on the applicability of single-parameter-scaling in the 
crossover region.
In section IV, we calculate the second and third cumulants of the conductance distribution
and analyze their $1/\langle g\rangle$ corrections. In the last section we summarize our findings.

\section{Model}

We have studied numerically the zero temperature conductance of the
2D Anderson model, described by the Hamiltonian
\begin{eqnarray}
H = \sum_{i} \epsilon_{i}a_{i}^{\dagger}a_{i}+ t\sum_{i,j}
a_{j}^{\dagger} a_{i}+ {\rm h.c.} \;, \label{hamil}
\end{eqnarray}
where the operator $a_{i}^{\dagger}$ ($a_{i}$) creates (destroys)
an electron at site $i$ of an square lattice and $\epsilon_{i}$ is
the energy of this site chosen randomly between $(-W/2, W/2)$ with
uniform probability. The double sum runs over nearest neighbors.
The hopping matrix element $t$ is taken equal to $-1$, which set
the energy scale, and the lattice constant equal to 1, setting the
length scale.  We have considered square samples  of size
$L\times L$. All calculations are done at an energy
equal to $-1$, to  avoid
possible specific effects associated with the center of the band.

The zero temperature conductance $g$   is proportional to the transmission
coefficient $T$ between two semi--infinite leads attached at the opposite
sides of the sample,
\begin{equation}
g= \frac{2e^2}{h}T,
\label{res}\end{equation}
where the factor of 2 comes from spin. From now on, we will measure
the conductance in units of $2e^2/h$.
We have calculated the transmission coefficient from the Green function,
which was obtained propagating layer by layer with the recursive
Green function method \cite{M85}.
This drastically reduced the computational effort, and we can easily solve
samples with lateral dimension up to $250$ for the calculation of the
distribution function, which
requires a huge number of independent runs to get good statistics
in the tails.
The number of different realizations employed is of $10^6$
for most values of the parameters.
We have considered wide leads with the same section as the samples, which
are represented by the same hamiltonian as the system, Eq.\ (\ref{hamil}),
but without diagonal disorder.
The leads serve to obtain the conductivity
from the transmission formula in a way well controlled
theoretically and close to the experimental situation.
In the study of the distribution function, we use cyclic periodic boundary
conditions in the direction perpendicular to the leads. In the calculation
of the corrections to the variance and the skewness, we have also considered
hard wall boundary conditions to make sure that they do not drastically change
the results. The main conclusions are similar and we will present results for
periodic boundary conditions, for which we have better statistics.

\section{Conductance distribution}

\begin{figure}[b]
\includegraphics[width=.48\textwidth]{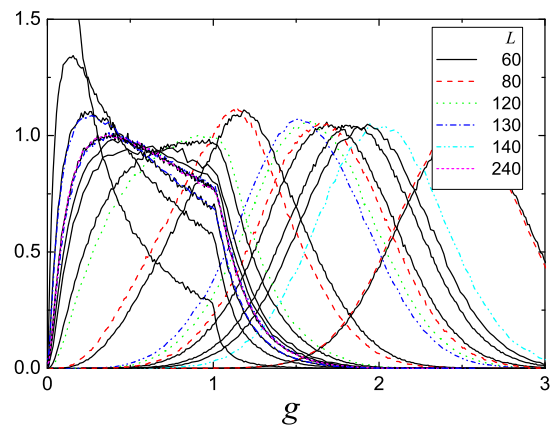}
\caption{(Color online) Conductance distribution function in the crossover between the diffusive
and the localized regime regime for the values of $L$ size indicated in the legend and for $W$ ranging between 3 and 6.}
\label{fig1}
\end{figure}

We have obtained the conductance distribution function for many
values of the disorder and the system size, chosen in such a way that the system is in
the crossover region, i.e., its mean value of the conductance is close to one.
In figure 1 we represent many of these distributions as a function of the conductance for the values of the system size given in the legend. The disorder varies between $W=3$ and 6.
When the mean conductance is larger than one the distribution is basically a gaussian
with an approximately constant standard deviation, given by the value of the universal conductance fluctuations.
As we approach the crossover, the shape of the high conductance tail remains close
to a gaussian function, but there is a drastic change
of behavior of the distribution at the  value $g=1$.
An important conclusion that can be extracted from Fig.\ 1 is that the entire
conductance distribution is uniquely determined by the mean conductance along
the crossover region. This
support the strong version of the single parameter scaling hypothesis. In several cases,
we have adjusted the value of the disorder in such a way that the mean conductance is the
same  for  different system sizes. In all these cases, the agreement between
the different distribution functions is quite remarkable. For example, the distribution
function for $W=4.2$ and $L=60$, whose mean conductance is equal 0.6677, is very similar to the distribution
for  $W=4.75$ and $L=240$, whose mean is 0.6686.

For some distributions, we can appreciate a marked discontinuity in the first derivative
of the distribution at $g\approx 1$.
This drastic change of behavior has been reported in different contexts,
but it has never been analysed in depth for 2D systems. It appeared in the critical conductance distribution
at the metal-insulator transition in 3D systems\cite{Ma06},
at the quantum Hall transition \cite{KO05}, in the 2D symplectic ensemble \cite{Ma94} and in
quasi-one-dimensional systems \cite{GS01,GM02}. In order to check if this apparent discontinuity in the
fist derivative of $P(g)$ is a finite size effect
or not, we have run extensively three different systems with the same mean conductance,
equal to $0.67\pm 0.01$, and with sizes equal to 60, 120 and 240 (the values of the disorder are 4.75, 4.43 
and 4.2 respectively).
In Fig.\ 2 we plot the
conductance distribution functions of these three systems. We first note the excellent
agreement between the three of them, showing again that the distribution is enterily determined by
a single parameter. Second, we appreciate that the possible
discontinuity in the first derivative of the distribution does not seem to vanish away
as the size increases.

\begin{figure}[htb]
\includegraphics[width=.48\textwidth]{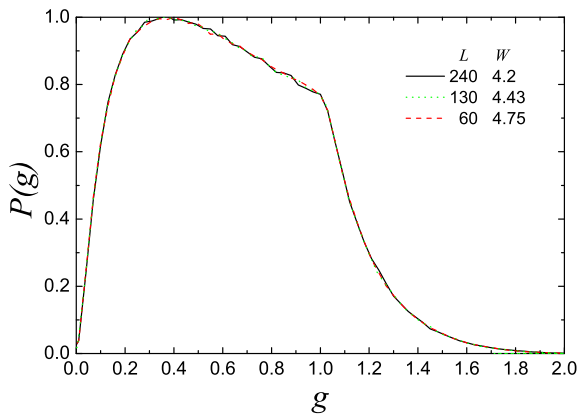}
\caption{(Color online) Conductance distribution function for three different pairs of values of the disorder and the
system size chosen so that the mean conductance is the same for all of them, equal to $0.67\pm 0.01$.}
\label{fig2}
\end{figure}

\section{$1/g$ contributions to the skewness}

The large amount of data gathered in our simulation allows us to study with enough precission how the conductance distribution deviates from a gaussian in the diffusive regime
as a result of $1/g$ corrections.
We have quantify this deviation through the skewness, Sk, of the conductance.
As we expect a leading contribution to Sk proportional to $1/\langle g\rangle$, we have plotted
in Fig.\ 3 Sk$\langle g\rangle$ as a function of $1/\langle g\rangle$ for periodic
boundary conditions.
The data corresponding to the ballistic regime have not been drawn. The criteria to
consider a data in the ballistic regime was the ratio of system size to elastic mean free path, $L/l$, to be smaller than 20. We will discuss this criteria later on when analyzing the behavior
of the variance. The elastic mean free path can be obtained through the dependence of the mean conductance with the system shape  \cite{beenakker}. For our model, we define $l$ through the equation
\begin{equation}
\langle g\rangle=\frac{L_xl}{L_y}-\frac{1}{3}
\end{equation}
where $L_x$ and $L_y$ are the transversal and longitudinal dimensions, respectively.
We have fitted the data in Fig.\ 3 by an expression of the form
\begin{equation}
\mbox{Sk}\langle g\rangle=a+\frac{b}{\langle g\rangle}+\frac{c}{L}
\label{fit}\end{equation}
where the last term corresponds to a finite size contribution.
The straight line in Fig.\ 3 is the extrapolation to infinite size of
Eq.\ (\ref{fit}) and intersects the vertical axis at $-0.0005\pm 0.0004$.
This value is in reasonable agreement with the prediction
of van Rossum {\it et al.}\cite{vRLAN} of $-0.002$ , if we take into account that
this estimate does not considered an specific set of boundary conditions
and so we are not comparing exactly the same quantities.
We note that in the region studied, the contribution to the skewness proportional
to $1/\langle g\rangle^2$ is large and soon dominates over the negative linear contribution.

\begin{figure}[tb]
\includegraphics[width=.48\textwidth]{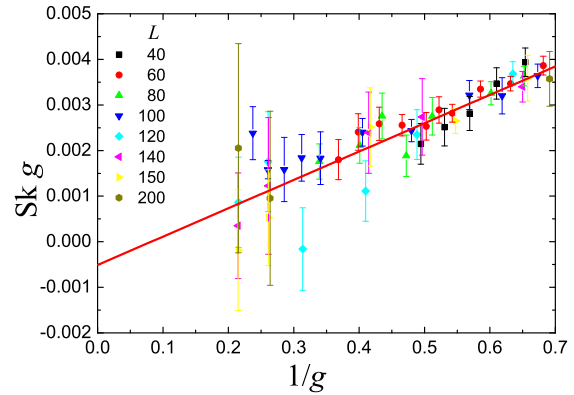}
\caption{(Color online) Skewness multiplied by $\langle g\rangle$ as a function of $1/\langle g\rangle$ for periodic  boundary conditions. Each symbol corresponds
to a different system size.}
\label{fig3}
\end{figure}

We have also calculated the variance of the conductance, $\sigma^2=\langle g^2\rangle_c$. In Fig. 4 we plot the variance of the data that we consider to be in the diffusive regime as a function of the inverse of the mean conductivity $1/\langle g\rangle$.
The horizontal line corresponds
to the value of universal conductance fluctuations for periodic
boundary conditions, which is equal \cite{markos} $\sigma_0^2=0.1544$. We note that the range of
the vertical axis is small.
There is a region in the ballistic regime where the variance presents a peak (not shown), whose magnitude depends on system size and is larger than the value
of the universal conductance fluctuations \cite{markos}. 
In Figs. 3 and 4 we have only included data with $L/l>20$, our criteria to be in the diffusive regime. We have calculated the eigenvalues of the transmission matrix, and for  $L/l<20$ there are many with transmission close to one. i.e., there are many ballistic
channels contributing to the current. As $g$ increases the variance
gets closer to the universal value.

\begin{figure}[tb]
\includegraphics[width=.48\textwidth]{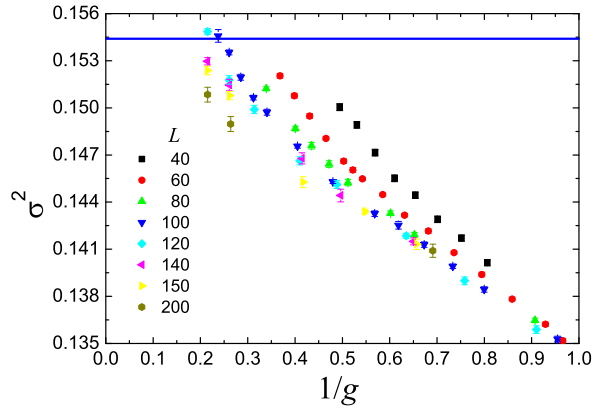}
\caption{(Color online) Variance as a function of $1/\langle g\rangle$ for periodic
 boundary conditions. Each symbol corresponds
to a different system size. The horizontal line represents the value of the universal conductance
fluctuations.}
\label{fig4}
\end{figure}

We note in Fig.\ 4 that there is a systematic size dependence of the variance,
breaking the strong version of single parameter scaling in this regime.
This is probably due to the ballistic regime, whose influence is much larger than expected.
The existence of these finite size effects together with the strong criteria needed to be in the diffusive regime, makes difficult a precise numerical determination of the universal conductance fluctuations.

\section{Discussion}

We have found numerically the behavior of the variance and the skewness in the crossover regime is compatible with analytical predictions \cite{AKL86}. However,  our results do not constitute a clear check since more precise numerical study of $1/\langle g\rangle$ corrections to the variance requires, on one hand, a very large number of realizations and on the other
hand rather large sizes in order to avoid the ballistic regime. 

We have also observed the change in shape of the distribution function when from diffusive through crossover regime. It can hardly be attributed, though, to the existence of  prelocalized states \cite{AKL86,FE95} which cannot be numerically checked for available system sizes at present time. This implies that the pre-exponential factor for these states should be relatively small.

\acknowledgements
We acknowledge financial support from the Spanish DGI (FIS2009-13483), Fundacion Seneca (08832/PI/08 and 11602/EE2/09) and  EPSRC (T23725).


\end{document}